\documentclass[12pt]{iopart}
\usepackage{verbatim}
\usepackage{epsfig}
\usepackage{amssymb}
\usepackage{floatflt} 
\usepackage{caption}
\usepackage{graphicx}

\begin{document}

{\title[$R_{AA}$ of Electrons from
heavy-flavour decays in Pb-Pb collisions at
\mbox{$\sqrt{s_{\rm NN}}=2.76$ TeV} ~~]{\large Measurement of the nuclear modification factor of electrons from
heavy-flavour decays at mid-rapidity in Pb-Pb collisions at
\mbox{$\sqrt{s_{\rm NN}}$ = 2.76 TeV} with ALICE}}
\vspace{-0.1cm}
\author{Yvonne Pachmayer$^1$ for the ALICE Collaboration}

\address{$^1$ Physikalisches Institut der Universit\"at Heidelberg,  \mbox{Philosophenweg 12, 69120 Heidelberg}, Germany}
\ead{pachmay@physi.uni-heidelberg.de}
\vspace{-0.1cm}
\begin{abstract}
We present results on inclusive electrons for 1.5  $ < p_{\rm T} < $ 6 GeV/$c$ in \mbox{Pb-Pb} collisions at $\sqrt{s_{\rm NN}}$ = 2.76 TeV measured with ALICE at the LHC and compare these to a cocktail of background electron sources. The excess of electrons beyond the cocktail at high momenta (\mbox{$p_{\rm T} >$ 3.5 GeV/$c$}) is attributed to electrons from heavy-flavour decays. The corresponding nuclear modification factor indicates heavy-flavour suppression by a factor \mbox{of 1.5-4}. 
\end{abstract}
\vspace{-0.8cm}
\vspace{-0.1cm}
\section{Introduction}\vspace{-0.3cm}
A Large Ion Collider Experiment (ALICE) \cite{ALICEExperiment} is the dedicated heavy-ion experiment at the LHC. It is
assumed that in central nucleus-nucleus collisions at high energy a Quark-Gluon Plasma (QGP), a high density deconfined state of strongly interacting matter, is formed. 
Heavy-flavour quarks (c, b) are excellent probes of the QGP. Originating from initial scattering processes they are produced on a very short time scale ($\tau$ $\approx$ 1/(2$m_{\rm q}$) $\lesssim$ 0.1 fm/$c$). Thus they are sensitive to the full history of the collision. In particular, they enable us to study parton energy loss and its quark mass dependence. \newline
\noindent Heavy-flavour production is measured in several channels by ALICE. We present first results of indirect measurements of charm and beauty quarks in Pb-Pb collisions at $\sqrt{s_{\rm NN}}=2.76$ TeV via the identification of single electrons: c or b $\rightarrow$ ${\rm e}^{\pm}$ + X \mbox{(B.R. $\simeq$ 10\%).}
\vspace{-0.1cm}
\section{Electrons from heavy-flavour decays}
\begin{figure}[htbp]
     \begin{minipage}{0.4\textwidth}
       \includegraphics[width=1.2\textwidth]{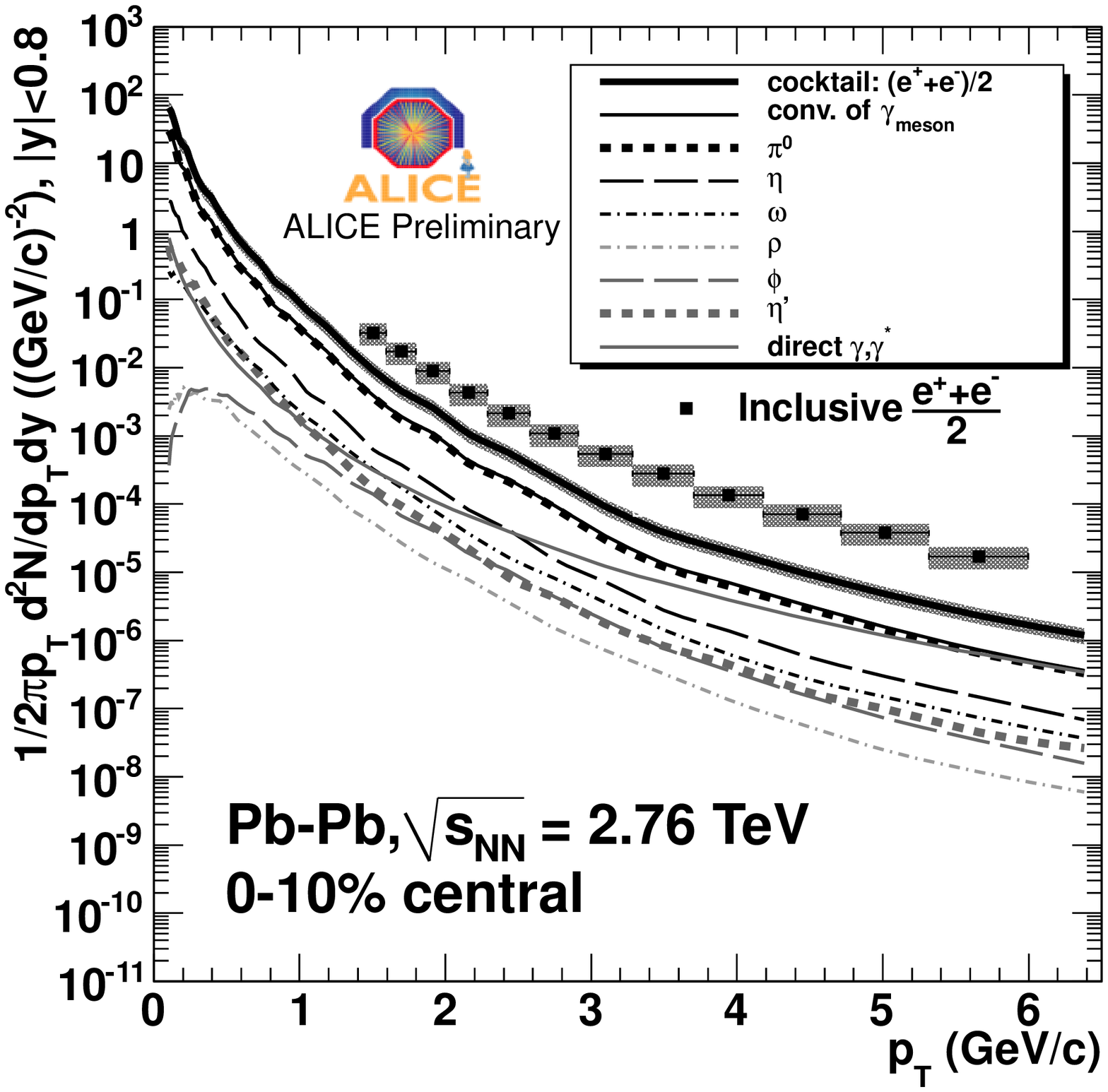} \label{figure2}
       \vspace{-1.2cm}
       \captionsetup{singlelinecheck=off}
       \caption{\small Inclusive electron spectrum for the centrality \mbox{0-10\%} compared to the cocktail of background electron sources.}
     \end{minipage}
     \hspace{2.3cm}
     \begin{minipage}{0.4\textwidth}
       \includegraphics[width=1.2\textwidth]{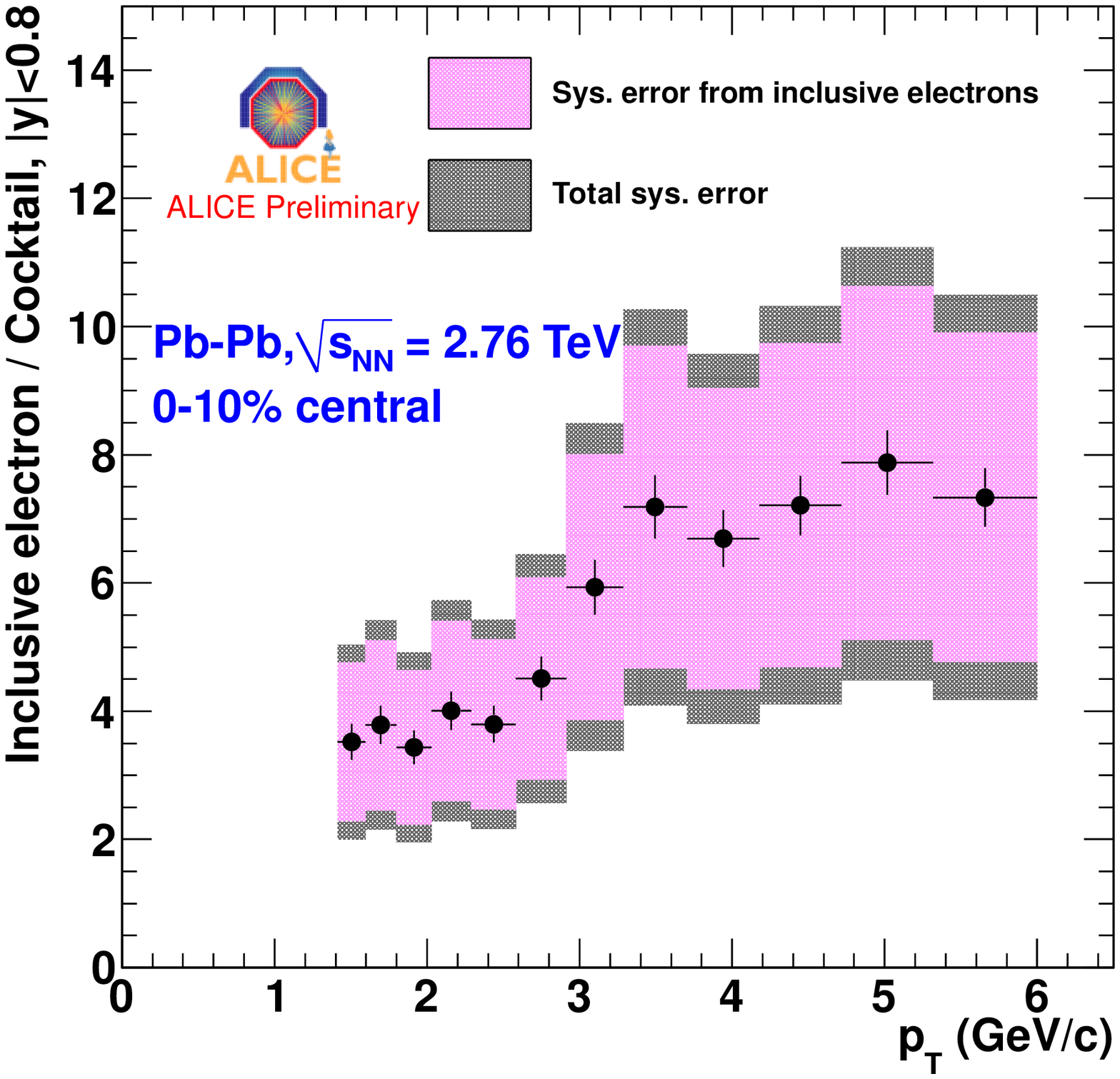} 
       \vspace{-1.2cm}
       \captionsetup{singlelinecheck=off}
       \caption{\small Ratio of inclusive electron spectrum to cocktail of background electron sources for the centrality \mbox{0-10\%.}}\label{figure3}
     \end{minipage}
   \end{figure}

\vspace{-0.3cm}
The results presented in this contribution are obtained from the first Pb-Pb run at \mbox{$\sqrt{s_{\rm NN}}=2.76$ TeV}. The Silicon Pixel Detector (SPD) at mid-rapidity and the forward VZERO scintillator counters provide a minimum bias trigger (corresponding to 97\% of the inelastic Pb-Pb cross section), and are also used to derive the centrality of the \mbox{collisions \cite{AToia}}. The ALICE detector system allows excellent track reconstruction and particle identification for electrons in the central barrel ($|\eta|<0.9$). In the present analysis the Inner Tracking System (ITS) and the Time Projection Chamber (TPC) were used for vertex and track reconstruction. After selecting good-quality tracks in the TPC, we required a hit in the innermost layer of the ITS (r = 3.9 cm) to reduce the background from photon conversions. The measured time of flight had to be consistent with the electron hypothesis within 3$\sigma$ to suppress kaons up to $p$ = 1.5 GeV/$c$ and protons up to \mbox{$p$ = 3 GeV/$c$}. Furthermore, a cut on the specific energy deposit d$E$/d$x$ in the TPC expressed in number of sigmas from the electron line was applied. Electrons were selected within the range 0 to 3 $\sigma_{\rm e}$ to reduce pions as much as possible. The remaining hadron contamination was determined via fits of the TPC d$E$/d$x$ in momentum slices, and the yield was subtracted from the electron spectra. In the momentum range \mbox{1.5-6 GeV/$c$} the hadron contamination amounts to less \mbox{than 10\%}.
\begin{floatingfigure}[l]{0.5\linewidth}
\vspace{-0.6cm}
\begin{center}
\includegraphics[width=0.5\linewidth,height=0.5\linewidth]{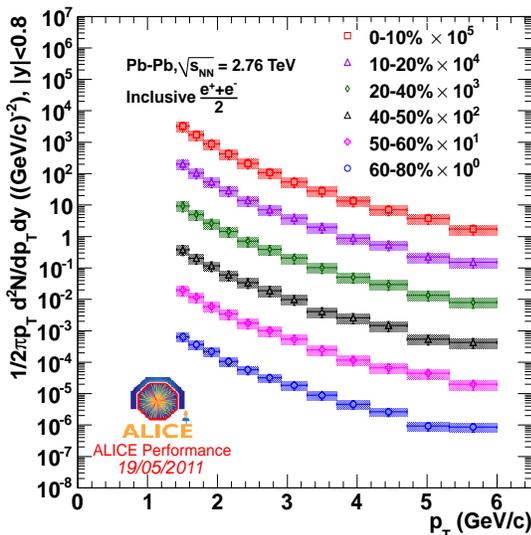}\end{center}\vspace{-0.4cm}\captionsetup{singlelinecheck=off,justification=RaggedRight}\vspace{-0.5cm}\caption{\small Efficiency and acceptance corrected $p_{T}$ spectra of inclusive electrons for various centrality ranges.}\label{figure1}
\end{floatingfigure}
\noindent The efficiency and acceptance corrections were derived from Monte Carlo (MC) simulations. The TOF matching efficiency was cross-checked with a data-driven method in which the signal from \mbox{$\rm \gamma\rightarrow$ e$^{+}$e$^{-}$} conversions in material was evaluated. The resulting transverse momentum spectra for inclusive electrons for various centrality ranges are shown in \mbox{Fig. \ref{figure1}}. The systematic uncertainty of 35\% is mainly given by the uncertainties in the particle identification. There is ongoing work to extend the $p_{\rm T}$ range towards low $p_{\rm T}$. In the future the information of the Transition Radiation Detector (TRD) and the Electromagnetic Calorimeter (EMCAL) will also be used in the analysis. This will improve the particle identification. Thus the systematic uncertainty will significantly be reduced and the spectrum can be extended towards \mbox{higher $p_{\rm T}$}. \newline
\noindent The inclusive electron yield consists primarily of three components: (i) electrons from heavy-flavour hadron decays, (ii) photonic background electrons from Dalitz decays of light mesons ($\rm \pi^{0}$, $\rm \eta$, etc) and photon conversions as well as real and virtual direct photons and (iii) non-photonic background from dielectron decays of vector mesons and $K_{\rm e3}$ decays. The photonic background (ii) is the largest background contribution. To extract the electrons from heavy-flavour hadron decays the background sources are subtracted with the so-called "cocktail method". Using a MC event generator of hadron decays a cocktail of electron spectra of background sources is calculated based on the yield and the momentum distributions of the electron sources. Figure \ref{figure2} shows the inclusive electron spectrum for 0-10\% most central collisions and the corresponding cocktail with the presently included background sources. Electrons from $\rm \pi^{0}$-Dalitz decays \mbox{($\rm \pi^{0} \rightarrow$ $\rm \gamma$ e$^{\rm +}$e$^{\rm -}$)} and from photon conversions from the decay ($\rm \pi^{0} \rightarrow \rm \gamma \rm \gamma$) in the detector material contribute dominantly to the background. Thus an excellent parametrization of the pion spectrum is mandatory. At the moment our pion input is based on charged pion measurements with ALICE \cite{PionInput}. The heavier mesons are implemented via $m_{\rm T}$ scaling. The QCD photons (direct $\rm \gamma, \rm \gamma^{*}$) are based on NLO calculations \cite{WVogelsang}. The systematic uncertainty is conservatively estimated to be 25\%. In the future the systematic uncertainty of the cocktail can be decreased with an improved pion input and direct measurements of other background sources. \newline
\vspace{-0.6cm}
\vspace{-0.4cm}

\noindent Figure \ref{figure3} shows the ratio of the inclusive electron spectrum to the cocktail of background electron sources for the centrality 0-10\%. We attribute the yield beyond the cocktail in the $p_{\rm T}$ region 3.5-6 GeV/$c$ to semileptonic heavy-flavour decays. In addition there is a hint of an excess at small transverse momentum, which depends on centrality. It is observed that for peripheral events the ratio is compatible with the corresponding result in pp collisions at 7 TeV. This could point to an additional electron source at low $p_{\rm T}$, e.g. thermal radiation. 

\vspace{-0.1cm}
\section{Nuclear modification factor}\vspace{-0.3cm}
Information on the properties of the Quark-Gluon Plasma can be obtained by comparing a given observable in nucleus-nucleus collisions ("QCD medium") to the one measured in pp collisions ("no QCD medium"). Among the most interesting observables is the suppression of high $p_{\rm T}$ hadrons quantified with the nuclear modification factor:
$R_{\rm AA}(p_{\rm T}) = \frac{1}{ \langle T_{\rm AA} \rangle} \cdot \frac{dN_{\rm AA}/dp_{\rm T}}{d\sigma_{\rm pp}/dp_{\rm T}}$,
where  $\langle T_{\rm AA} \rangle$ is the average nuclear overlap function for a given centrality range, $dN_{\rm AA}/dp_{\rm T}$ and $d\sigma_{\rm pp}/dp_{\rm T}$ represent the particle yield in nucleus-nucleus collisions and the cross section in pp collisions, respectively. 
\noindent \mbox{Figure \ref{figure4}} shows the nuclear modification factor of background-subtracted electrons (see above) for the centrality ranges 0-10\% (central events, $\langle N_{\rm part} \rangle$=357) and 60-80\% (peripheral events, $\langle N_{\rm part} \rangle$=23).
\noindent The pp reference spectrum is obtained by scaling the measured spectrum of electrons from heavy-flavour decays in pp at \mbox{$\sqrt{s}$ = 7 TeV} \cite{SMas} to 2.76 TeV based on
\begin{floatingfigure}[l]{0.5\linewidth}
 \vspace{-0.6cm}
\begin{center}
\includegraphics[width=0.5\linewidth,height=0.5\linewidth]{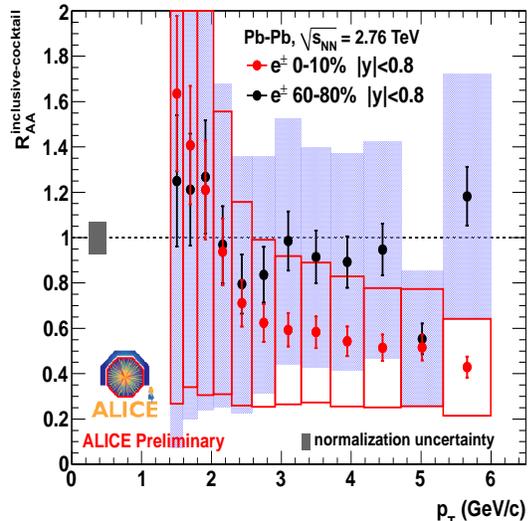}\end{center}\vspace{-0.4cm}\captionsetup{singlelinecheck=off,justification=RaggedRight}
\vspace{-0.5cm}\caption{\small Comparison of the $R_{\rm AA}$ of background subtracted electrons for central and peripheral Pb-Pb collisions}\label{figure4}
\end{floatingfigure} 
\noindent FONLL calculations \cite{ExtrapolationPaper}. The systematic uncertainties in the FONLL calculation (i.e. uncertainties of the bare quark masses, the scale parameters and differences in scaling for electrons from charm and beauty decays) are $\sim$10\% for \mbox{$p_{\rm T} >$ 2 GeV/$c$}. In contrast to peripheral events, a suppression of a factor of \mbox{1.5-4} is found for 0-10\% central collisions in the $p_{\rm T}$ region 3.5-6 GeV/$c$, where charm and beauty decays dominate. This indicates strong coupling of heavy quarks to an opaque medium, that is created in heavy-ion collisions. The result is compatible with the $R_{\rm AA}$ of muons from heavy-flavour \mbox{decays \cite{XCheu}.} 

\begin{flushleft}
\vspace{-0.3cm}
\section{Conclusion and outlook}\vspace{-0.6cm}
\end{flushleft}
We have measured inclusive electron spectra for 1.5  $ < p_{\rm T} < $ 6 GeV/$c$ in \mbox{Pb-Pb} collisions at $\sqrt{s_{\rm NN}}$ = 2.76 TeV. The electron yield beyond the expected background cocktail at high momenta \mbox{($p_{\rm T} > $ 3.5 GeV/$c$)} is attributed to electrons from heavy-flavour decays. The corresponding nuclear modification factor for 0-10\% most central collisions shows a suppression of a factor of 1.5-4, which indicates strong coupling to the medium but is nevertheless higher than the $R_{\rm AA}$ of D mesons (charm only) \cite{ADainese}. The suppression tends to vanish towards peripheral events. In addition a hint of a centrality dependent excess above background sources is observed at small momenta \mbox{($p_{\rm T} < $ 3 GeV/$c$)}. A cross-check with the D mesons measurement in ALICE at mid-rapidity seems to exclude an additional charm source at low $p_{\rm T}$. Thermal radiation could be another electron source. In the near future, the $p_{\rm T}$ range of the $R_{\rm AA}$ will be extended towards low and high $p_{\rm T}$. The systematic uncertainty will be reduced with improved particle identification, cocktail input and the pp reference spectrum measured in pp collisions at \mbox{$\sqrt{s}$ = 2.76 TeV}. Finally, the charm and beauty contributions will be disentangled on the basis of electron separation from the interaction vertex.

\end{document}